# Short story of stratospheric and ground-based observations of solar photosphere with high angular resolution in the 70s of the XX century at the Pulkovo Observatory

L.D.Parfinenko

Central (Pulkovo) Astronomical Observatory, St. Petersburg, Russia

**Abstract**

In the 70s of the last century, the stratospheric solar observatory "Saturn" with a 100cm telescope was launched at Pulkovo Observatory. The photographs and spectra obtained on it for more than 45 years remained record-breaking in angular resolution. Then, for ground-based observations of Sun fine structure in Pamir at an altitude of 4.5 km, a 50 cm open-type Pulkovo mobile telescope was installed. Photographs and spectra obtained with the Saturn telescope for the visible range were an important step in the development of high spatial resolution technologies in heliophysics. However, to date, the details of these studies remain unknown to the global scientific community. In this paper, an attempt is made to partially fill this gap.

Studies of fine structure of the solar atmosphere are related to solving the most important problems of solar physics,

The inhomogeneities in the brightness of the visible solar surface were discovered by James Short in 1748, and the term "granule" was introduced by Daves in 1864. The first photographs of solar granulation were received by Pierre Jansen in 1877. Temporary sequences of photographs of granulation and spots were obtained by A.P. Gansky in Pulkovo in 1905-1907.

The first studies of the fine structure of the solar atmosphere were based on optical observations with a high spatial resolution of the order of 0.5 "(~ 400 km on the Sun). Already the first observers, having received photographs of the solar surface, realized the distorting influence of the earth's atmosphere. Therefore, in 1874–75, attempts to raise an observer with a small telescope on balloons to a height of 6000–8000 m above the ground began in France. One of the flights ended in the tragic death of two aeronauts and after that the balloon observations were interrupted for a long time. Balloon flights resumed again in France in the middle of the twentieth century by Dolphus brothers (Charles is the pilot, and Andonin is the observer astronomer) together with D.E. Blackweell. They photographed (November 22, 1956 and April 1, 1957) solar granulation from a height of 6000 m with a refractor with a lens diameter of 29 cm and in the last flight in 4 hours received 480 images and many with a resolution of about 0.8 "(~ 600 km), and the observed granulation contrast K reached 4.8%.

In the 50s, thanks to technological progress, it was already possible to use automatic equipment on board the balloon, since the presence of a person on it is associated with many difficulties and has become completely unjustified. Moreover, it became possible to raise the stratospheric balloon to a height of more than 20,000 m.

In 1955, M. Schwarzschild and colleagues at Princeton University Observatory began the development of a stratospheric solar telescope with a main mirror diameter of 30.5 cm in order to obtain direct images of the surface of the Sun with a diffraction resolution 0.45"( Stratoscope – 1). The use of automatic system and a thin (50 μm) polymer film for the balloon shell made it possible to conduct observations from previously inaccessible heights of 25 km, where the air density is 30 times lower than the ground.

The "spectrostratoscope" (D = 320 mm) created in Germany (Kipenhoyer, Meltretter, Opton) was similar to the "Stratoscope-1", but besides direct shots in the "white" light, it was planned to obtain spectrograms and photographs with Hα - filter. The dispersion of the spectrograph reached 10 mm / Å, and the spectral resolution of 12 mÅ. In this case, the shutter speed was 10 s, which made great demands on the stability of the image position. For the first time, the telescope was suspended from a balloon by its center of gravity and a three-stage image stabilization system was used.

There was only one successful flight on May 17, 1975. In 10 hours, at an altitude of 27 km, 400 spectrograms and more than 1000 "white" direct images were obtained. The resolution on them changed little and approximately corresponded to the theoretical limit of the telescope aperture (R = 0.37 ", K = 6%). The quality estimates of spectrograms of the Spectrostratoscope and any work done on their basis are unknown.

In the USSR, efforts were focused on creating a large stratospheric solar telescope for studying the atmosphere of Sun with previously unattainable spatial resolution. In May 1961, a special resolution of the government of the USSR was issued on the creation of a stratospheric solar telescope with a mirror lens diameter of 50 cm, and then 100 cm. For this work until the end of the 70s. last century allocated a lot of funding. This was a project in which hundreds of specialists from several factories and scientific institutes participated. The scientific management of the project was carried out by Pulkovo Observatory (Krat et al., 1972). The main work on the creation of the telescope was successfully carried out by the Kazan Optical and Mechanical Plant. The main designer of the project was V.F. Mikhalev.

The most convenient means of lifting the telescope is an automatic aerostat, which can lift a load weighing several tons to a height of 20-30 km and about a ton to a height of 40 km. In this case, there are no many serious difficulties arising from observations from space or when using fast-moving vehicles heavier than air.

Distortions limiting spatial resolution during observations from the Earth arise in the atmosphere at altitudes of less than 15 km. Above, the air density is low and there are practically no distortions. Therefore, to achieve a high resolution of ~ 0.1 ", the telescope does not have to be carried out into airless space. It can be in the atmosphere, but above the distorting layers. In this case, the telescope itself should not introduce instrumental distortions.

Start, rise and flight in this case do not require energy, can be quite smooth, without overloads. The balloon moves with the surrounding air at relatively low speeds (usually less than 100 km / h) and the airflow around the telescope suspended from it is negligible. The small distance from the launch site greatly simplifies communication with the stratospheric observatory and receiving telemetric information.

After the observations, the aerostat shell was shot off, and all the equipment, along with the obtained observational materials, gently landed by parachute. In this case, there were practically no damages and, thus, conditions were created for repeated use of the equipment with a minimum amount of work on its repair and preparation for the next start.

In addition to the possibility of increasing spatial resolution in the stratosphere, the spectral range extends 100 times. The spectral range containing many important solar ultraviolet and infrared lines, for example, D$\alpha$ 6561.03, He 10830, bands of the CO type, practically becomes free from atmospheric absorption. In the stratosphere, atmospheric scattered light and restrictions due to cloudiness are practically absent. At the same time, the cost of stratospheric research is 1-2 orders of magnitude less than space. When observing at wavelengths $\lambda \sim 3000$Å, the usual photospheric granulation in the center of the disk begins to be replaced by facular granules. In the UV range around $\lambda \sim 2000$Å, conventional granulation is no longer visible. A two-dimensional picture of the power spectrum of the intensity of images obtained in the continuum reveals a large number of peaks (modes) that can noticeably change their frequency from place to place on the surface of the Sun. The most stable and pronounced scale of the fine structure of the photosphere is a granular grid of 1.5-2.4 arc seconds (1 "~ 725 km).

However, the conditions under which the stratospheric observatory operates are very harsh. After 2 hours of lifting, the equipment passes from ground temperature, in the summer about + 20. ° C, to the stratospheric –50º ÷ -60 °C. Convective heat sink is reduced, heating by radiation from the Sun and the Earth is increased, temperature differences are created up to 100°.

In a short time, the Soviet Stratospheric Solar Observatory, codenamed "Saturn", was created. The telescope has a main mirror from a special optical glass (SITALL) with a diameter of 1 meter to obtain from a height of 20 km images and spectra of solar structures with a resolution of up to 0.12 ″ (this corresponds to 90 km on a sun surface) unattainable at that time from the earth's surface due to atmospheric distortions. The most successful was the third flight of telescope "Saturn" in 1970 (the main mirror was used with a diameter of 0.5 m), when 93 frames with a quality corresponding to the theoretical resolution for a 50-cm mirror were obtained. Spectrograms with a resolution of ~ 0.5 ″ were also obtained.

In the 4th flight of 1973 with a 1-meter main mirror, the resolution of several images was brought to 0.12″. For comparison, in 1985 in the images of the Space Shuttle (USA) it was only 0.4″. Before the creation of solar telescopes with adaptive optics (the beginning of XX century), photographs and spectrograms of the Sun obtained with telescope "Saturn" remained unsurpassed (for the visible range) in angular resolution and calibration quality. The best negatives are digitized and many scientific studies have been carried out on their basis.

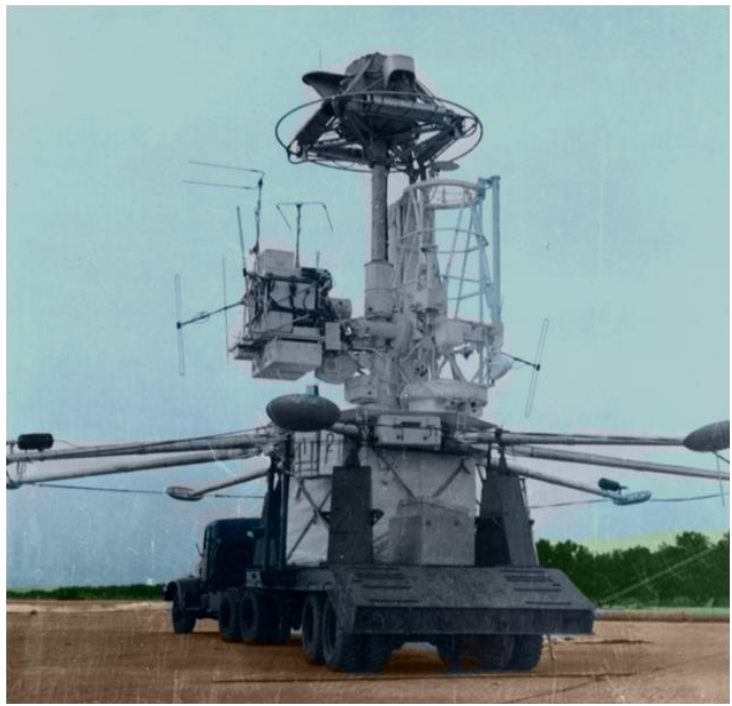

Figure 1 Stratospheric observatoty "Saturn" at the start of the last IV flight of 1973 at the Volsk training ground (near the Saratov city)

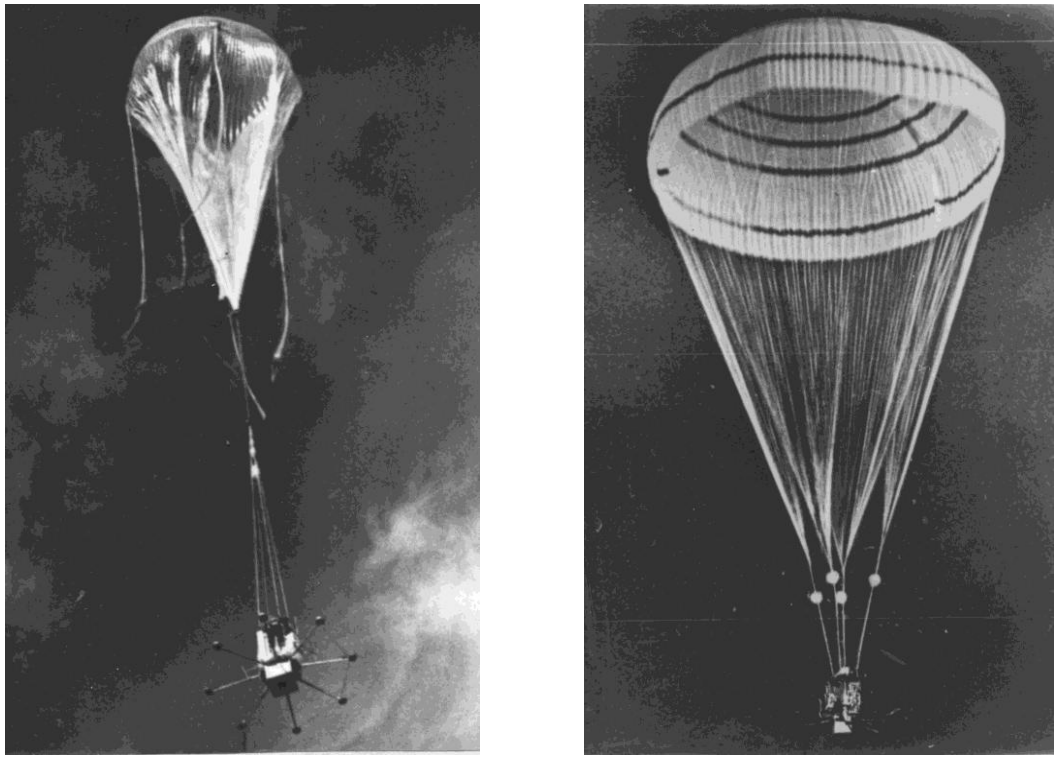

Figure 2 Stratospheric observatoty "Saturn": left -in flight, right- parachute landing

Fig.3. shows one of 25 top-quality shots of photosphere 2 '× 3' in size. It was obtained on June 30, 1970 at 8h44m10s (UT) at "Saturn" Pulkovo stratospheric observatoty. The exposure time is 1ms, the flight altitude is 21 km. (stratospheric balloon-borne observatory, Krat et al., 1970). The image scale is 1 mm = 1 ".72. Using a color filter, a wavelength band of 4380-4800A (effective wavelength of 4600A) was selected. An 80 mm FT30 film calibrated by a 9-step attenuator was used. This minimized distortion of the granulation contrast. The telescope had an alt azimuth mount. The optical scheme of cassegrene is used. The parabolic mirror of the telescope and additional hyperbolic are aluminized and coated with a protective layer of silicon oxide.

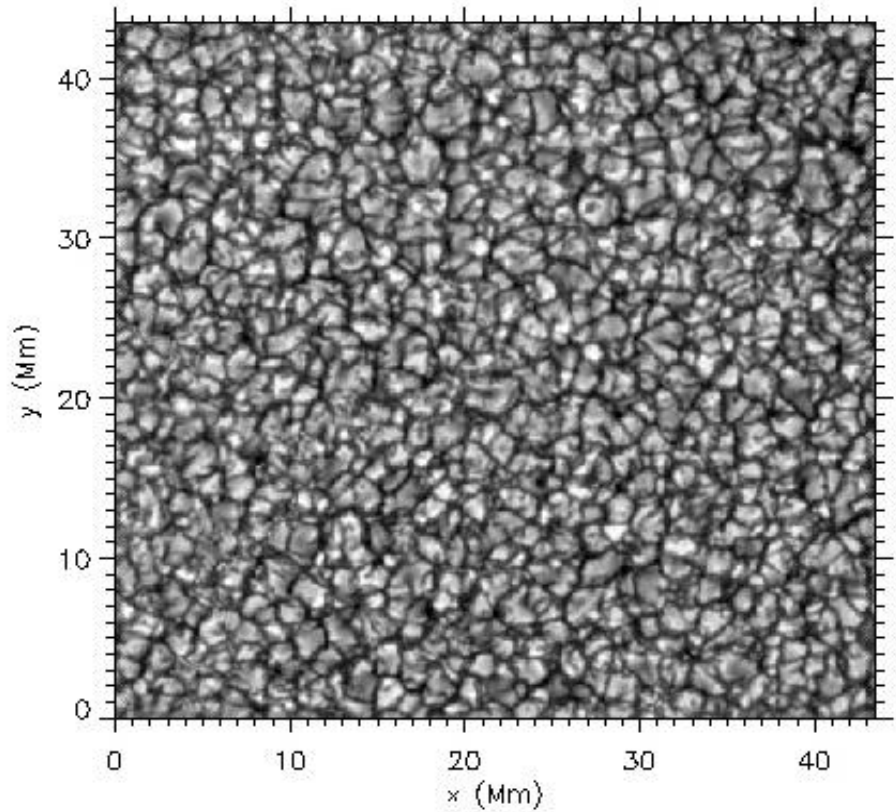

Figure 3 Photo of solar granulation, 3rd flight of the stratospheric observatory "Saturn" 1970

The light diameter of the parabola is 50 cm, the focal length is 120 m. The mirror is made of SITALL - an artificial polycrystalline material with high uniformity and a linear expansion coefficient of zero in the temperature range from −60 ° C to + 120 ° C. The achieved angular resolution is 0 ".24, which corresponds to the diffraction resolution of 50 cm of the telescope objective. In 1973, an improved version of "Saturn" stratospheric telescope with a mirror diameter of 100cm was launched. The diffraction resolution of the lens 0 ".12 (Krat, 1974), was achieved only in several photo of the photosphere. Even today, after more than 47 years, this is a very good resolution. A photoelectric image quality analyzer measuring the sharpness of granulation was used. For this, scanning along a circle with a frequency 55 Hz of a special prism, a photometric section of the solar granulation in the frame is created, and then the signal is passed through a narrow-band electric filter, rectified and smoothed. When the maximum values are reached, the shutters of the direct cameras and spectrograph are triggered. We used a two-stage guidance system, automatic focusing of the image, an opening heat-shielding curtain, etc. The telescope was directed to the desired part of the Sun from the earth by an astronomer using a television system. The photographs and spectra of Sun obtained at Pulkovo stratospheric telescope "Saturn" remained record until June 2009, when 100cm German-Spanish stratospheric solar telescope (for the UV range) the "Sunrise" (Barthol et al. 2010), was launched. It has the best technical characteristics and is equipped with a digital vector magnetograph. The main advantage of balloon telescopes over space observatories is the order of magnitude lower project cost and the availability of modernization. During a 137-hour flight in the polar region of Canada, the "Sunrise" telescope obtained valuable scientific material on the fine structure of the magnetic field.

The observational material obtained at the "Saturn" stratospheric solar observatory has outstripped the existing world level in terms of quality, resolution, and reliability of photometric calibration. Despite the launch of the stratospheric 100 cm (for the UV range) telescope "Sunrise", the images and spectra obtained with the "Saturn" telescope for the visible range are remain an important stage in the development of observational methods of heliophysics.

For further systematic observations, a substantial modernization of the entire complex of the stratospheric telescope was required. The management of Pulkovo Observatory decided to create a more advanced version - the stratospheric station "ISKRA» (scientific supervisor Karpinsky V.N.). This is a high quality 40cm telescope reflector for the ultraviolet range. It applied new technical solutions, for example, introduced the 4th stage of image guiding and provided for the use of a computer. All technical documentation was created and the main components of the telescope were made. At the end of the 80s, a successful test flight of the telescope to a height of 40 km was completed. But unfortunately, due to the closure of astronomical topics at the Kazan Optical and Mechanical Plant, work at the "ISKRA" station was completely discontinued, as well as all other scientific stratospheric projects and developments in the USSR.

Stratospheric astronomy remains significantly cheaper than space methods, so the development of new telescopes for carrying to high altitudes up to 50 km continues. There are serious projects to create stabilized platforms on tethered balloons, which constantly hang over a specific area at an altitude of about 10 km.

Unfortunately, stratospheric observations do not allow a systematic study of the fine structure of solar atmosphere. Therefore, using the technical developments of the Saturn stratospheric station, the Pulkovo Observatory and the Kazan Optical and Mechanical Plant developed and installed in 1978 in the eastern Pamir at an altitude of 4300 m near the city of Murgab an open-type ground-based solar telescope with short optical paths (project supervisor Parfinenko L .D.) The exceptionally dry local climate allowed the use a telescope without a pavilion. It is probably the world's largest mobile ground-based solar telescope for study of fine structure. The Pamir solar telescope is partially made on the basis of mechanics and optics of the stratospheric telescope "Saturn" remaining from a 50cm variant, replaced by a 100cm telescope "Saturn", which made a successful flight in 1973.

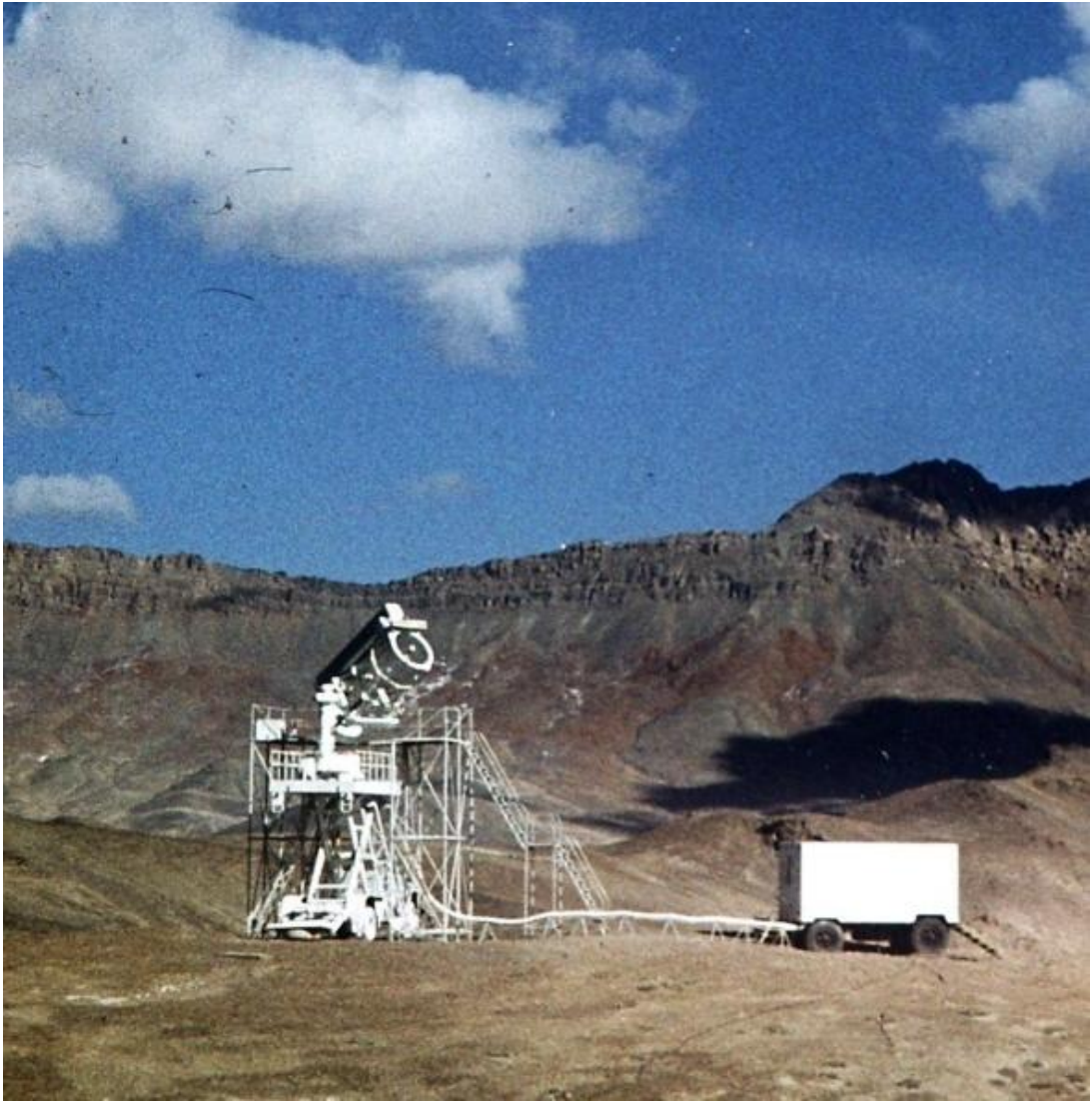

Figure 4 Pamir solar telescope near the expedition base of Pulkovo Observatory. The village of Shorbulak (37 ° 55 '30.39 "N, 74 ° 9' 17.88" E). Eastern Pamir, Gorno-Badakhshan Autonomous Region of Tajikistan. The height of the telescope is 4350 m above sea level. Most of the electronic equipment, including the telescope control panel with a television sight monitor, is located 12 meters east of the telescope in a warmed trailer - an operator.

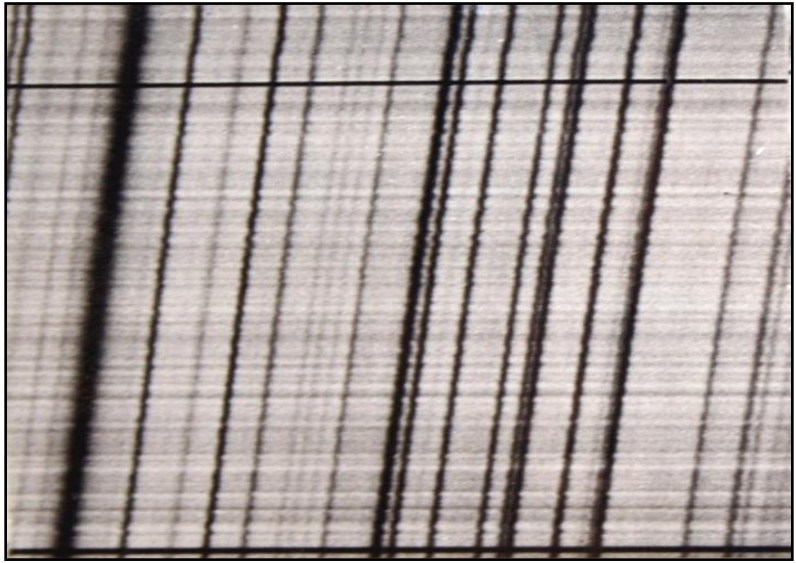
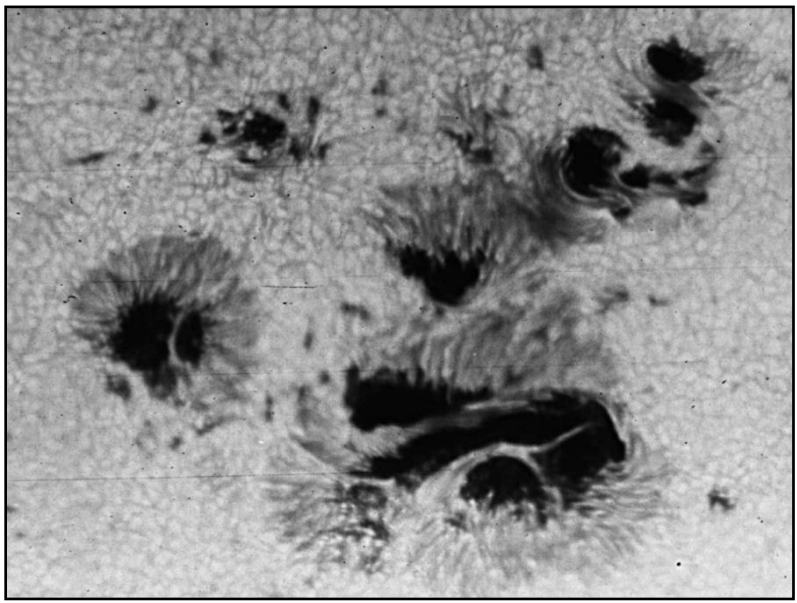

Figure 5  Fragment of the spectrogram obtained on the Pamir telescope and a photography of a group of sunspots (a 6-hour series of observations obtained on the Pamir telescope on September 22, 1979)

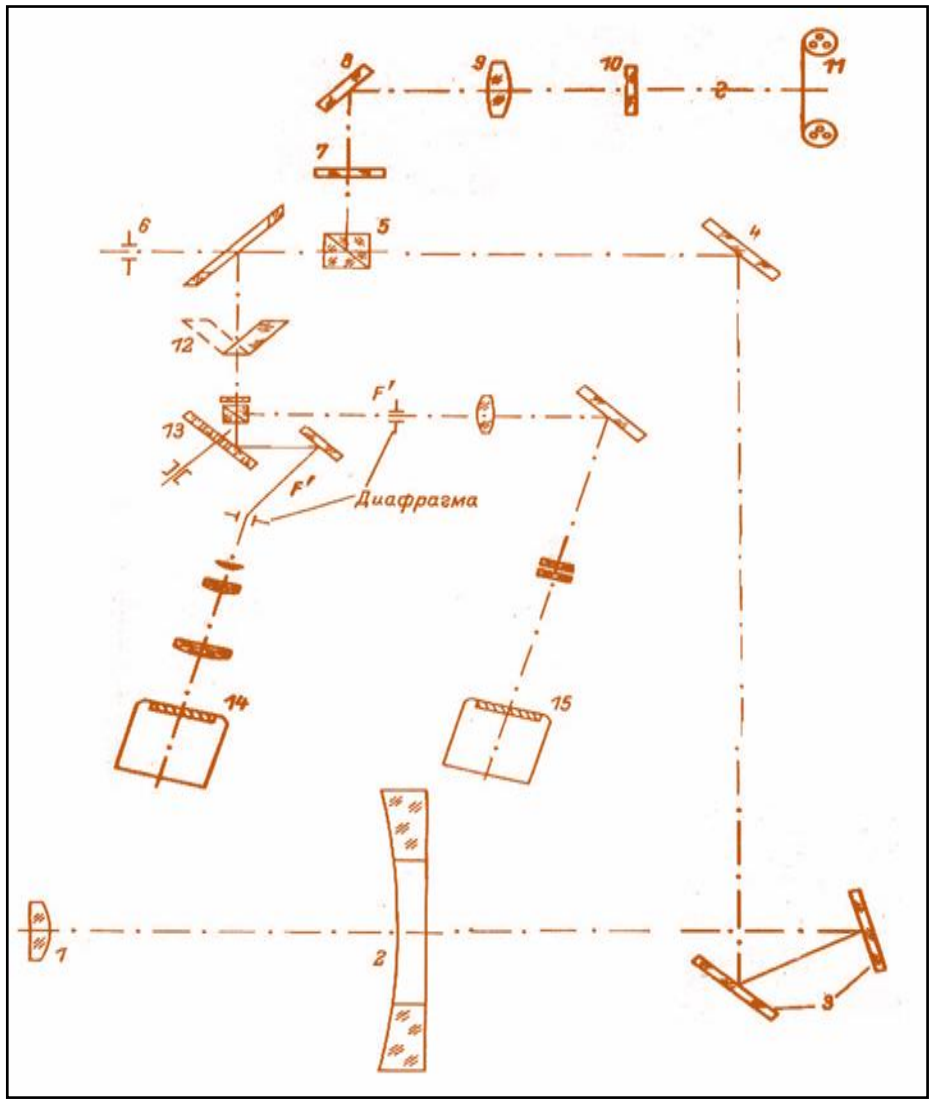

Figure 6 Simplified optical scheme of Pamir solar telescope:

    1, 2 - hyperbola and parabolic mirror,

    3 - two-mirror auto focus unit,

    4 - additional mirror,

    5 - beam splitting cube,

    6 - slit spectrograph,

    7,8,9,10,11 - channel camera direct photos,

    12 - rotating prism,

    13- modulator of the quality analyzer and automatic focusing system,

    14.15 - photomultiplier.

Spectrograph Pamir telescope was made according to the Ebert-Fasti scheme with crossed dispersion, which made it possible to accommodate three sections of the spectrum of different orders located one above the other in one frame. The grill has three different positions, so that nine areas in the range 3900-6600A are photographed in total. The widths of the sections vary from 100A to 200A, and the dispersion from 1A / mm to 0.5A / mm. The specular slit of the spectrograph is made oblique, which made it possible to take images of the image on the slit in spectral line Hα using the Halle IPF. Aerial film type -17 of 19 cm format was used as photographic material in the camera of the spectrograph. In the camera of direct shots, FT-30 technical film of 8 cm format was used.

The technique of "hungry" manifestation of long rolls of FT-30 film and wide aerial film of type 17 was used. Special drum developing machines for 120 meters of film were made, in which the film pre-soaked with discified water was repeatedly scrolled through photo-solutions. Laboratory studies showed good uniformity of manifestation along the entire length of the roll, not different from the usual cuvette manifestation, and the absence of the Eberhard effect. Films are calibrated from the solar beam by step attenuators. Special prisms double the illumination on the step attenuator, expanding the light range.

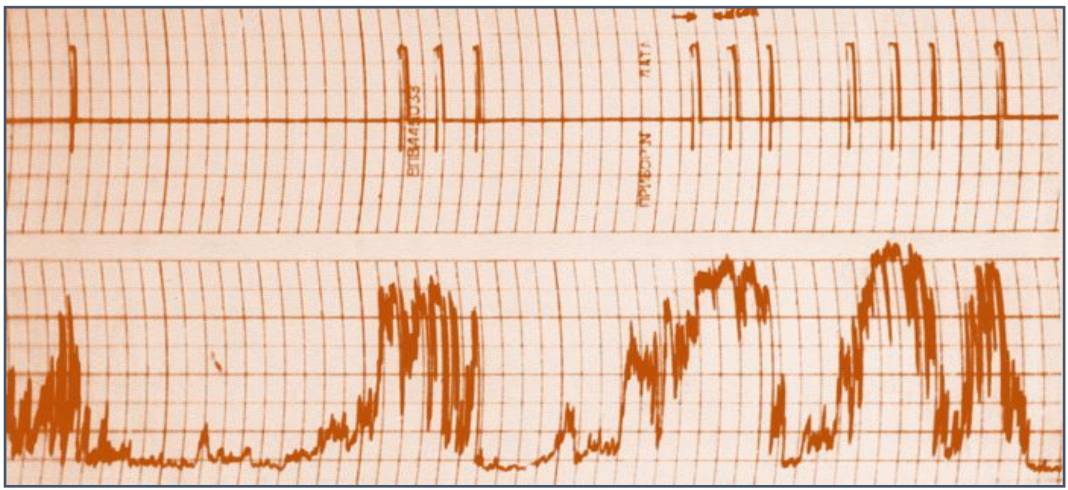

Figure 7 Fragment of the recording on the tape of the recorder of the readings of the image quality analyzer for harmonic 1.5″ (1 time division 5 sec.). Above, the shutter release timestamps are visible.

After aiming at the target object on the Sun with a television sight, the operator sets a certain threshold for the operation of the automatic quality analyzer and does not interfere with the operation of the telescope. The telescope automatically monitors (there is a two-stage photo guide) the entire object throughout the daylight hours and takes pictures of the image of the Sun and spectra at times when the signal exceeded the set threshold of the quality analyzer. The telescope is equipped with a special programmer that allows you to work fully automatically without an astronomer operator. The Pamir solar telescope is actually the world's first prototype of a large solar telescope - a robot. The Pamir telescope is mounted on the lifting platform of a special car. For transportation, the platform is lowered and alt-azimuthal system of the telescope is laid on the car in a horizontal position using a hydraulic jack. Transferring the telescope from the traveling position to the working one takes about an hour. In Shorbulak the telescope is installed in a stationary version. It is one of the first robotic telescopes operating without the participation of an operator-atronomer based on signals from an image quality analyzer. For 11 years of work (1978-1989) at the Pamir telescope, the author and his colleagues received unique scientific material on the fine structure of solar atmosphere. Unfortunately, due to the well-known events in Afghanistan and the subsequent collapse of the USSR, the Pulkovo Observatory lost access to this unique telescope.

Over the past decade, the use of new technologies has led to a revolution in the quality of the resulting terrestrial images of Sun. Currently, several ground-based solar telescopes equipped with adaptive optics systems allow systematic high-resolution observations. They are equipped with large light mirrors made using modern technologies. Such are, for example, the 1.6 meter GST reflector (http://www.bbso.njit.edu) equipped with adaptive optics and image speckle reconstruction, the New Vacuum Solar Telescope (NVST) in China (http://fso.ynao.ac .cn / index.aspx) and others. These ground-based telescopes provide spatial resolution better than that achieved from the stratosphere and even from space for a given cost of the telescope.

At present, the idea of open-type automatic telescopes with short optical paths has received a rebirth after the advent a digital systems registration image, solid-state light detectors and adaptive optics systems with powerful processors, as well as speckle interferometry methods. For example, such is the open-type telescope DOT (Netherlands) (https://web.archive.org/web/20021016231735/http://dot.astro.uu.nl/)

Conclusion

The observational material obtained at the Saturn stratospheric solar observatory in quality, resolution and reliability of photometric calibration is significantly ahead of the world level that existed at that time. Photographs and spectra of the Sun obtained with telescope "Saturn" for the visible range remain an important technological achievement in modern heliophysics. Only in 2009, the more advanced 100cm German-Spanish stratospheric solar telescope (for the UV range) "SUNRISE" (Barthol et al. 2010) was launched.

Currently, former 100cm stratospheric telescope "Saturn" has been converted to ground-based and installed at the Pulkovo Observatory. The telescope has been modernized; digital drives and a modern light receiver (CCD matrix) are installed on it. It turned out to be advisable to use it to solve some astrometric problems: asteroids, exoplanets, etc. (http://www.gaoran.ru/english/as/p2015/bikulova.pdf)